\documentclass[12pt]{iopart}

\newcommand{\ket}[1]{|#1\rangle}
\newcommand{\bra}[1]{\langle#1|}

\usepackage{graphics}
\usepackage{hyperref}
\begin{document}

\title[Photon pair generation using four-wave mixing in a microstructured fibre]{Photon pair generation using four-wave mixing in a microstructured fibre: theory versus experiment}

\author{O. Alibart$^1$, J. Fulconis$^1$, G. K. L. Wong$^2$, S.G. Murdoch$^2$, W. J. Wadsworth$^3$ and J. G. Rarity$^1$}

\address{$^1$Centre for Communications Research, Department of Electrical and Electronic Engineering\\
University of Bristol, Merchant Venturers Building, Woodland Road, Bristol, BS8 1UB, UK}

\address{$^2$Department of Physics, University of Auckland, Private Bag 92019, Auckland, NZ}

\address{$^3$Photonics \& Photonic Materials Group, Department of Physics\\
University of Bath, Claverton Down, Bath, BA2 7AY, UK}

\ead{olivier.alibart@bristol.ac.uk}

\begin{abstract}
We develop a theoretical analysis of four-wave mixing used to generate photon pairs useful for quantum information processing. The analysis applies to a single mode microstructured fibre pumped by an ultra-short coherent pulse in the normal dispersion region. Given the values of the optical propagation constant inside the fibre, we can estimate the created number of photon pairs per pulse, their central wavelength and their respective bandwidth. We use the experimental results from a picosecond source of correlated photon pairs using a micro-structured fibre to validate the model. The fibre is pumped in the normal dispersion regime at $708\,nm$ and phase matching is satisfied for widely spaced parametric wavelengths of $586\,nm$ and $894\,nm$. We measure the number of photons per pulse using a loss-independent coincidence scheme and compare the results with the theoretical expectation. We show a good agreement between the theoretical expectations and the experimental results for various fibre lengths and pump powers.
\end{abstract}

\maketitle

\section{Introduction\label{intro}}

Quantum information can be encoded in various media including atoms, ions and electrons. It is photons however that are the most useful for transporting quantum information between separate locations. They are often called \emph{flying Q-bits} and they appear to be the fundamental resource for quantum communications experiments~\cite{QKDwein, QKDzbin}. Their robustness to decoherence also leads them to be involved in various multiphoton and linear optical logic experiments~\cite{cnotrarity,expCNOT,Qordi}. One of the main resources needed for these quantum optics experiments is the ability to create these photons in pairs. When simply correlated in time they can be used to create heralded single photons~\cite{pdcrarity,pdcmandel} (closely approximating true single photon sources). More interesting is when these photons have strongly correlated properties; they are entangled~\cite{brightpdc1,brightpdc2}. Such sources continue to be exploited in fundamental multiphoton experiments such as quantum teleportation~\cite{teleporthugues} or to create cluster states~\cite{ghz,4phot,5phot}.

The preferred sources for such experiments until recently have been three-wave mixing in $\chi^{(2)}$ non-linear birefringent crystals~\cite{brightpdc1,brightpdc2}. These sources are inherently wide band, low brightness (per nanometer, per single mode) sources. The increasing number of photons involved in current experiments is requiring brighter sources and narrower photon bandwidths. Hence, periodically poled fibres~\cite{pdcfiber2} and periodically poled waveguides of lithium niobate~\cite{Tanzeleclett} have been shown to be bright pair photon sources. In poled fibres~\cite{brightpdc3} the low non-linearity limits the brightness while in planar waveguides the non-circular mode limits the coupling efficiency into optical fibres~\cite{moi}.

It is well known that parametric gain can arise from the $\chi^{(3)}$ non-linearity in optical fibres~\cite{agrawal,wang} and phase matching can be achieved by using the modulation instability when pumping fibres in their anomalous dispersion regime. Various pair photon generation experiments have been performed in this regime~\cite{PCFkum0,PCFkum1,PCFkum,PCFkum2,PCFsagnac}. The photon pairs are generated close to the pump wavelength and are always accompanied by a significant Raman background and careful filtering is required. We have shown that phase matching can be obtained for widely spaced wavelengths by pumping photonic crystal fibre (PCF) close to the zero dispersion wavelength in the normal dispersion regime~\cite{PCFUKcleo,PCFbath,PCFUK2,gordon0}. Recently, we have used a fibre with the zero dispersion point in the near infra-red ($715\,nm$) and pumped with a picosecond pulsed laser, red detuned a few nanometers into the normal dispersion regime, to demonstrate a high brightness source of photons pairs~\cite{PCFUK}.

We are seeking to develop a source which may be applicable for future quantum interference experiments involving three or more photons created as two or more pairs. Interference effects between separate pairs of photons can be studied by overlapping photons with a time uncertainty shorter than their inverse bandwidth or coherence length~\cite{fundprob}. This restricts us to sources pumped by ultra-short laser pulses where the bandwidth is of order nanometers and also requires a high efficiency of collection. Theoretical models of correlated photon pair generation via four-wave mixing have already been investigated in the CW approximation~\cite{wang}, but this approximation no longer stands if we want these photons ready for quantum interference experiments. We propose here the first quantum model to describe four-wave mixing to generate photon pairs in the pulsed regime in a single mode fibre.

In this paper, we investigate the generation of correlated single-photon pairs in photonic crystal fibres from both theoretical and experimental points of view. We first describe the four-wave mixing process in the quantum mechanical regime. We then analyze and compare the experimental results from correlated photon pairs generated in a photonic crystal fibre to the aforementioned theory. The paper is structured along the following lines:  \sref{FWM} presents the quantum theory of four-wave mixing in optical fibres, while \sref{thexp} is dedicated at tagging experimental parameters to the theory. \sref{exp} describes and analyzes the experimental results from our photonic crystal fibre. We investigate in this section the agreement between theoretical expectations and experimental results. \sref{disc} outlines the possibilities of using this fibre as a source of multi-photon pairs for quantum information processing. Finally, we briefly conclude in \sref{CCL}

\section{Four-wave mixing process in fibre with pulsed pumping\label{FWM}}
\subsection{Definition of the involved fields}
We are looking at the process of four wave mixing (FWM) based on the non-linear susceptibility $\chi^{(3)}$ in a photonic crystal fibre where two pump-photons are converted into a pair of signal and idler photons. We are specializing to energy matched situations where $2\Omega_p=\omega_s+\omega_i$. The related interaction Hamiltonian in a volume within the fibre is:
\begin{equation}\label{Hint1}
H_{int}=\int_VUdV
\end{equation}
and the energy density associated with four fields coupled in a third order non-linear medium is:
\begin{equation}\label{U}
U=\epsilon_0 \chi^{(3)}(\Omega_p,\omega_s,\omega_i)E_p^2 E_s E_i
\end{equation}
Here $\epsilon_0$ is the permittivity of free space allowing us to measure $\chi^{(3)}$ in units of $V^{-2}m^2$. In the fibre optic community, it is common to rather use the nonlinear refractive index $n_2$ defined as follow:
\begin{equation}
	n_2=\frac{3\;\chi^{(3)}}{4\;\epsilon_0cn_0^2}
\end{equation}
where $n_0$ is the refractive index of the fibre. 

For a nonstationary pumping pulse, the field amplitude can be decomposed into its Fourier components in the following form:
\begin{equation}
	E_p\left(x,y,z,t\right)=\int_{-\infty}^{+\infty}\,\mathcal{E}_p\left(\omega,x,y,z,t\right)d\omega
\end{equation}
In our guided configuration all the involved fields are propagating in the same direction, hence we insert $e_p(\vec{r})$ as the transverse spatial variation with normalisation $\int\left|e_p(\vec{r})\right|^2 drd\theta=1$ inside all our equation and carry out the calculation along the $Z$-axis which is the fibre axis. We assume that $E_p\left(x,y,z,t\right)$ is a strong classical gaussian pump beam of full-width half maximum bandwidth $\Delta\omega_p$ and central frequency $\Omega_p$. We focus on one of its Fourier component defined as follow:
\begin{equation}\label{Ep}
\mathcal{E}_p(\omega)=\frac{E_{p_o}}{2}\,G_p\left(\omega\right)\left(e^{-i[\left(\Omega_p+\omega\right)t-(k_p-\gamma P_p)z]}+e^{i\left[\left(\Omega_p+\omega\right)t-(k_p-\gamma P_p)z\right]}\right)e_p(\vec{r})
\end{equation}
where $\gamma P_p$ is the self-phase modulation coefficient acquired by the pump pulse as it propagates within the fibre; $G_p\left(\omega\right)=e^{-\frac{\omega^2\sigma^2}{2}}$ is the weight of the component $\Omega_p+\omega$ of the pulse and $\frac{1}{\sigma^2}$ the variance characterizing the pump pulse. The variance is linked to the bandwidth of the pulse by $\Delta\!\omega_p=\frac{2\sqrt{\ln(4)}}{\sigma}$. We define the energy density within a pulse so that (Parseval's theorem):
\begin{equation}
U=\epsilon_0\int_{-\infty}^{+\infty}\,\mathcal{E}_p(\omega)\mathcal{E}_p^*(\omega)d\omega=\epsilon_0\int_{-\infty}^{+\infty}\,\left|\frac{E_{p_o}}{2}\;e^{-\frac{\omega^2\sigma^2}{2}}\right|^2d\omega
\end{equation}
While the strong pump pulse remains classical, here the two-photon modes associated with $E_s$ and $E_i$ are quantized using the creation and annihilation operators and we use similar spatial terms $e_l(\vec{r})$ to describe their transverse variation with normalisation $\int\left|e_l(\vec{r})\right|^2 drd\theta=1$.
\begin{equation}\label{El}
E_l=\sqrt{\frac{\hbar\omega_l}{2\epsilon_l}}\frac{1}{\sqrt{\mathcal{L}}}\sum_{k_l}\left(a^\dag_le^{-ik_lz}-a_le^{ik_lz}\right)e_l(\vec{r})
\end{equation}
where $\mathcal{L}^3$ defines the quantization volume.

\subsection{Interaction Hamiltonian}

We calculate the interaction Hamiltonian for one monochromatic component of the pump pulse and will sum up over all the possible combinations between the components at a later stage. Starting with the interaction Hamiltonian for the parametric process and following the steps in \ref{hamiltonian}, we can easily derive the interaction Hamiltonian for two pump-components $\omega_{p1}$  and $\omega_{p2}$ as:
\begin{equation}\label{Hint}
\fl\mathcal{H}_{int}=SI\hbar\frac{1}{\mathcal{L}}\sum_{k_s,k_i}G_p\left(\omega_{p1}\right)G_p\left(\omega_{p2}\right)
\left[a_s^\dag a_i^\dag e^{-i\left(2\Omega_p+\omega_{p1}+\omega_{p2}\right)t}Lsinc\left(\frac{\Delta kL}{2}\right)
+h.c.\right]\nonumber
\end{equation}
where $I=\int\!\!\!\int e_{p1}(\vec{r})e_{p2}(\vec{r})e_s(\vec{r})e_i(\vec{r}) dr d\theta$ is a normalised factor; $\Delta k = k_{p1}+k_{p2}-k_s-k_i-2\gamma P_p$ is the phase matching coefficient and $S$ is the gain parameter defined as follow:
\begin{equation}\label{S}
S=\epsilon_0 \chi^{(3)}\frac{E_{p_o}^2}{4}\sqrt{\frac{\omega_s\omega_i}{4\epsilon_s\epsilon_i}}
\end{equation}
The sinc function in \eref{Hint} gives rise to the phase-matching condition between the pump components, signal and idler fields. The energy associated with the individual signal and idler fields in the fibre is defined by
\begin{equation}
H_o=\hbar [\omega_s(a^\dag_sa_s+1/2)+\omega_i(a^\dag_ia_i+1/2)]
\end{equation}
leading to a total Hamiltonian to be:
\begin{equation}
H_{tot}=H_o+\mathcal{H}_{int}
\end{equation}
Remembering that this Hamiltonian applies only for two arbitrary monochromatic pump components, we will later integrate over the full bandwidth of the pump pulse.

\subsection{The Heinsenberg equation of motion}

In the standard Heisenberg representation the evolution of any operator $A$ is defined by
\begin{equation}\label{heisen1}
\frac{dA}{dt}=-\frac{i}{\hbar}\left[A,H_{tot}\right]
\end{equation}
In the case of $A=a^\dag_s$ and assuming $a_l(t)=\tilde{a}_l(t)e^{-i\omega_lt}$, equation \ref{heisen1} gives:
\begin{equation}\label{as}
\fl\tilde{a}^\dag_s(t\rightarrow\infty)=\tilde{a}^\dag_s(0)
-i\frac{SIL}{2\sqrt{\mathcal{L}}}\;\sum_{k_i}G_p\left(\omega_{p1}\right)G_p\left(\omega_{p2}\right)\tilde{a}_i(0) \delta(\Delta\omega)sinc\left(\frac{\Delta kL}{2}\right)
\end{equation}
When integrating to obtain \eref{as} we assumed that the system (nonlinear medium and the pump components) is turned on at time zero and runs continuously. This gave rise to the $\delta$-function in $\Delta\omega$ for the frequencies of the fields to guarantee the energy conservation at single-photon level for a monochromatic pump component. 

We then perform the integration over all the possible combinations of $\omega_{p1},\omega_{p2}$ within the pump bandwidth. The full details of the following calculation are given in \ref{eqmotion} but basically if we define the interaction satisfying the basic energy conservation $2\Omega_p=\omega_{s_0}+\omega_{i_0}$ and introduce $\Delta\omega_{k_{s}}$ as the frequency shift for the signal mode $k_s$ compared to the central one $\omega_{s_0}$ (respectively $\Delta\omega_{k_{i}}$ for the idler mode $k_i$), the $\delta$-function then helps to carry out the double integral over $\left\{\omega_{p1},\omega_{p2}\right\}$  and for $\tilde{a}^\dag_s(\infty)$ we get 
\begin{equation}\label{asc}
\tilde{a}^\dag_s(\infty)=\tilde{a}^\dag_s(0)-i\frac{SIL}{2\sqrt{\mathcal{L}}} \frac{\sqrt{\pi}}{\sigma}\;
\sum_{k_i}e^{-\frac{\left[\Delta\omega_{k_s}+\Delta\omega_{k_i}\right]^2\sigma^2}{4}}\tilde{a}_i(0)sinc\left(\frac{\Delta kL}{2}\right)
\end{equation}
It is interesting to note that in \eref{asc} the dirac function (standing for the energy conservation in monochromatic pump regime) turns here into an exponential function guaranteeing the energy conservation in the broadband pump regime.

In order to calculate the mean number of photons $\left\langle N_s\right\rangle$ over all the possible modes $k_s$, we define the operator $W_s\left(t\right)=\frac{1}{\sqrt{\mathcal{L}}}\sum_{k_s}\tilde{a}_s(t)$ so that
\begin{eqnarray}\label{Ns}
\fl\left\langle N_s\right\rangle & = & _i\bra {0}_s\bra {0}W^\dag_sW_s\ket {0}_s\ket {0}_i\nonumber\\
\fl & = & \left(\frac{SIL}{2}\right)^2\left(\frac{\sqrt{\pi}}{\sigma}\right)^2\int\!\!\int sinc^2\!\left(\frac{\Delta kL}{2}\right)e^{-\frac{\left[\Delta\omega_{k_s}+\Delta\omega_{k_i}\right]^2\sigma^2}{2}}\frac{d\Delta\omega_{k_s}}{v_{g_s}}\frac{d\Delta\omega_{k_i}}{v_{g_i}}
\end{eqnarray}
where $v_{g_s}$ and $v_{g_i}$ are the group velocity of signal and idler wave in the fibre.

We show in the following section that \eref{Ns} becomes easy to integrate under certain approximations and we derive the expected mean number of photon per pulse at the output of our photonic crystal fibre.

\section{From theory to experiment\label{thexp}}
\begin{figure}[h]
\centerline{\scalebox{1.3}{\includegraphics{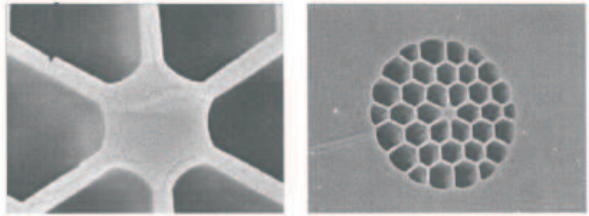}}}
	\caption{Electron microscope image of the PCF used with core diameter $d\approx2\,\mu\!m$ and $\lambda_0=715\,nm$.\label{pcf}}
\end{figure}

\begin{figure}[h]
	\centerline{\scalebox{1.1}{\includegraphics{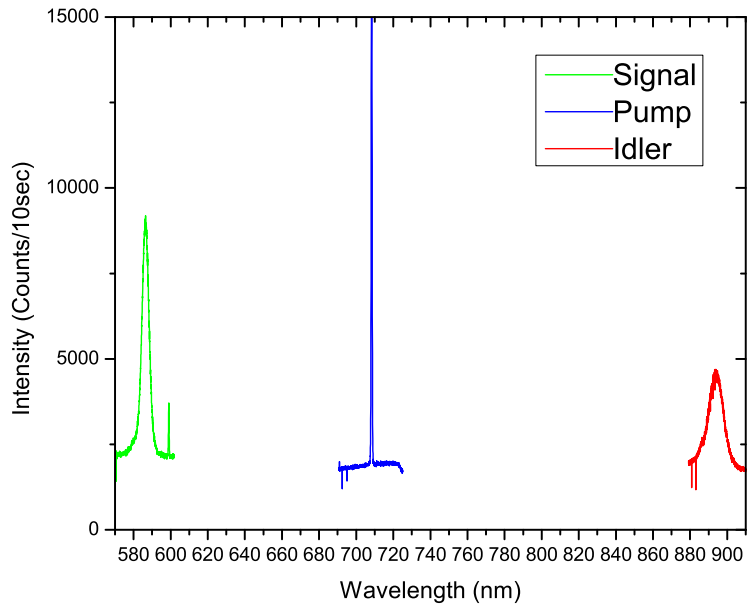}}}
	\caption{Fluorescence spectrum of the signal, pump and idler photons. The measurement was integrated over ten seconds. The number of counts is proportional to the number of photons detected by the cooled camera. The photon spectra are respectively centered at $586.4\,nm$, $708.4\,nm$ and $893.9\,nm$. They feature an associated FWHM bandwidth of $4.5\, nm$, $0.4\, nm$ and $9.6\, nm$. The small peak at $598\,nm$ is attributed to unwanted background light or a detector fault since it is visible even when the laser is blocked\label{sptotal}}
\end{figure}

\Fref{pcf} shows the microstructured fibre used in our experiment. The fibre has a core diameter of $2\,\mu\!m$ and presents a zero dispersion wavelength $\lambda_0=715\,nm$.
Typical spectra obtained when pumping with $708.4\,nm$ light and aligning the polarization on one of the axes of the fibre are shown in \fref{sptotal}. The asymmetry of the micro-structure makes the fibre slightly birefringent. As a consequence, when we do not launch into the correct fibre polarization axis, we see two peaks corresponding to different phase matching conditions for the different fibre axes. The absence of the second peak here implies our source is strongly polarised.

The short wavelength sits at $586.4\,nm$ and the corresponding idler at $893.9\,nm$. The spectra were taken at a pump power of $1\,mW$ and flat background comes entirely from the electronic bias errors in the CCD. Then by tuning the pump photon wavelength and monitoring similar spectra, we are able to observe the evolution of the signal and idler wavelength versus the pump wavelength (see \fref{expQPM}). These wavelengths are those which satisfy the following energy conservation:
\begin{equation}
	2\Omega_p-\omega_s-\omega_i=0
\end{equation}	
and phase matching equations
\begin{equation} 
	2\frac{n_p\,\Omega_p}{c}-\frac{n_s\,\omega_s}{c}-\frac{n_i\,\omega_i}{c}-2\gamma P=0
\end{equation}
where $n_{p,s,i}$ are the refractive index of the medium at pump, signal and idler wavelengths. The phase shift acquired by the pump waves (self-phase modulation) is defined by the factor $\gamma P$~\cite{agrawal}. We also present on \fref{expQPM}, the theoretical phase matched curve using the propagation constant for a simple strand of silica in air as a close approximation~\cite{corePCF} to the fibre used in this experiment. This approximation does not give a good agreement between the theoretical expectations and the experimental results as it does not take into account the role played by the honeycomb structure.

\subsection{Phase matched solutions\label{QAP}}
\begin{figure}[h]
	\centerline{\scalebox{1.1}{\includegraphics{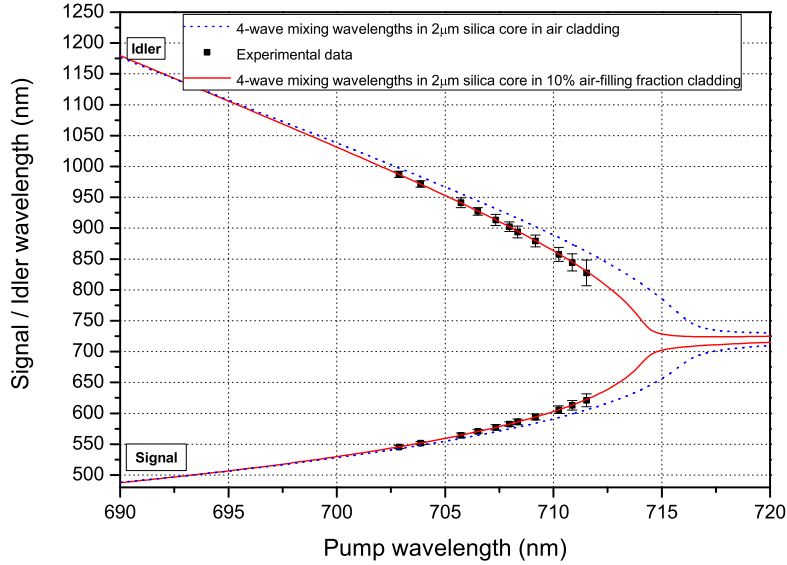}}}
	\caption{Nonlinear phase-matching diagram for the process $2\Omega_p\rightarrow\omega_s+\omega_i$. The dashed curve is the simulation using a plain strand of silica suspended in the air, while the straight curve is the fit used to retrieve the actual refractive index of the fibre. The black dots stand for the experimental data while their error bar is proportional to their measured FWHM.\label{expQPM}}
\end{figure}

As the percentage of silica in the cladding is not negligible, the accuracy of this model is not perfect. To solve this problem we use a novel method~\cite{PCFAUS} to determine the propagation constant of a photonic crystal fibre. This method models the PCF as a step-index fiber with a core index set to that of fused silica, and a cladding index set to the mean index of the PCF's cladding. We used an effective cladding index of $n=1.05$ (corresponding to $90\%$ air-filling fraction) to retrieve the propagation constant of the fibre and obtain the fit shown in \fref{expQPM}. There is a very good agreement, confirming that our core diameter is actually $2\,\mu\!m$ and that the honeycomb structure does contribute to the effective refractive index of the fibre cladding, perturbing the propagation constant from that generated by the naked strand model.

We have now a good knowledge of the propagation constant inside our fibre. This numerical data will allow us to simplify the calculation of \eref{Ns} and therefore predict the bandwidth and the mean number of photons per pulse.

\subsection{Number of photons per pulse and associated bandwidth\label{nbp}\label{bwdth}}

The integral from \eref{Ns} can be sorted out if we now develop the $\Delta k$-function around the phase matched frequencies ($2k_{po}-k_{so}-k_{io}-2\gamma P$=0) using a first order Taylor power series. 
\begin{eqnarray}
	\Delta k&=&	2k_p-k_s-k_i-2\gamma P\\
			&=& 2\left.\frac{\partial k_p}{\partial\omega}\right|_{\omega_p}\Delta\omega_p-\left.\frac{\partial k_s}{\partial\omega}\right|_{\omega_s}\Delta\omega_{k_s}-\left.\frac{\partial k_i}{\partial\omega}\right|_{\omega_i}\Delta\omega_{k_i}
\end{eqnarray}
With the help of energy conservation, we can remove the $\Delta\omega_p$ variable and find:
\begin{equation}\label{dk}
	\Delta k=\left(\mathcal{N}_s-\mathcal{N}_p\right)\Delta\omega_{k_s}+\left(\mathcal{N}_i-\mathcal{N}_p\right)\Delta\omega_{k_i}
\end{equation}
where $\mathcal{N}_l=\left[\omega_l\left.\frac{\partial n_l}{\partial\omega}\right|_{\omega_l}+n_l\right]$.

\begin{figure}[h]
	\centerline{\scalebox{0.75}{\includegraphics{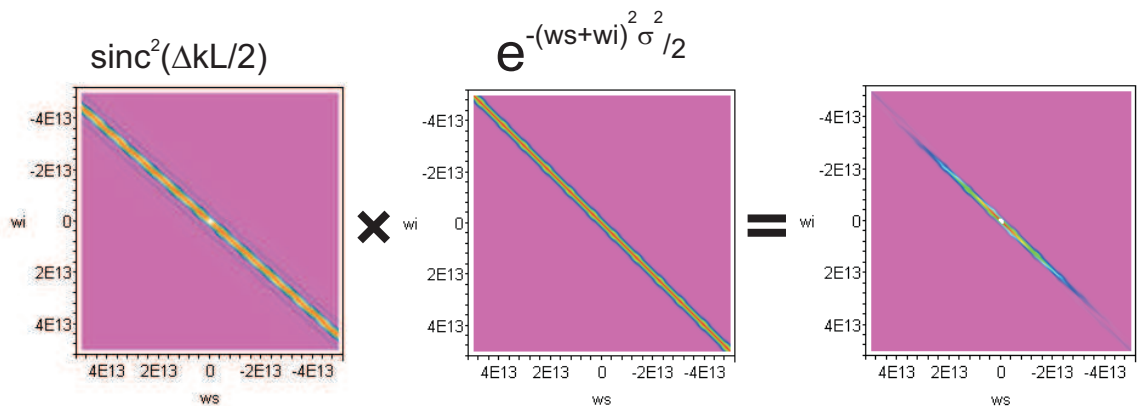}}}
	\caption{The hue color represents high values of the function. It is revealing to note the angle of the $sinc$ stripe which is controlled by the coefficients $\mathcal{N}_s,\mathcal{N}_i,\mathcal{N}_p$ --- arising from the fibre geometry. This is slightly different from the angle of the exponential stripe which is always fixed at $45^{\circ}$. The difference in angle between the two functions defines the narrowness of the signal and idler bandwidth.\label{trip}}
\end{figure}

We now have the products of two functions depending only on $\Delta\omega_{k_s}$ and $\Delta\omega_{k_i}$: 
The first one is $sinc^2\!\left(\frac{\Delta kL}{2}\right)$, standing for the natural phase matching condition in the PCF for a monochromatic pump. The second one is $e^{-\frac{\left[\Delta\omega_{k_s}+\Delta\omega_{k_i}\right]^2\sigma^2}{2}}$ standing for the pump pulse broadening. So if we want to sum over all the possibilities for our fibre we have to look at the product of these two functions and to integrate it over $\Delta\omega_s$ and $\Delta\omega_i$. We plot on \fref{trip} the graphical representation of both functions and their product. 
We can then carry out the formal calculation of $\left\langle N_s\right\rangle$ as explained in \ref{acalc} to eventually find:
\begin{equation}\label{Nsfinal}
	\left\langle N_s\right\rangle = \left(\frac{SIL}{2}\right)^2\left(\frac{\pi\Delta\omega_p^2}{4\ln(4)}\right)^{3/2}\left(\frac{4\sqrt{2}\pi c}{\left(\mathcal{N}_s-\mathcal{N}_i\right)L}\right)\frac{1}{v_{g_s}v_{g_i}}
\end{equation}

If we now develop the $\Delta k$-function around the phase matched frequencies using first order Taylor power series but use the energy conservation to remove the $\Delta\omega_{k_i}$ (respectively $\Delta\omega_{k_s}$) term, we get
\begin{eqnarray}
	\Delta\omega_{k_s}&=&\frac{2\pi c}{\left|\mathcal{N}_s-\mathcal{N}_i\right|L}+2\left|\frac{\mathcal{N}_i-\mathcal{N}_p}{\mathcal{N}_s-\mathcal{N}_i}\right|\Delta\omega_p\label{bdw}\\
		\Delta\omega_{k_i}&=&\frac{2\pi c}{\left|\mathcal{N}_i-\mathcal{N}_s\right|L}+2\left|\frac{\mathcal{N}_s-\mathcal{N}_p}{\mathcal{N}_i-\mathcal{N}_s}\right|\Delta\omega_p\label{bdw2}
\end{eqnarray}
giving respectively the bandwidth of the signal and idler photons. It is worth noting two separated terms: the first one corresponding to the natural bandwidth from a monochromatic pump, while the second one is the pump bandwidth broadening. In the classical CW case, even a perfectly monochromatic pump leads to down-converted photon with a finite bandwidth. This process behaves as the inverse of the length of the fibre~\cite{yariv}. Here, we have to take it into account and add the fact that our pump is not monochromatic. The pump bandwidth contribution is linked to the slope ($\left|\frac{\mathcal{N}_i-\mathcal{N}_p}{\mathcal{N}_s-\mathcal{N}_i}\right|$ for the signal and $\left|\frac{\mathcal{N}_s-\mathcal{N}_p}{\mathcal{N}_i-\mathcal{N}_s}\right|$ for the idler) of the phase matching curves in \fref{expQPM}.

\subsection{Experimental limitations\label{limit}}

In the previous treatment of the four-wave mixing process, we assumed that all the frequency components, and their following quantized operators, were all coherent within the pulse. Our calculation obviously applies to PCF pumped by Fourier transformed limited pulses, nevertheless it would be straightforward to decompose the pulse into several coherent subgroups, apply the treatment to and eventually sum up the number of photons coming from each of them.

Another experimental limitation we did not highlight in the previous treatment is group velocity walk-off between the three involved photons. Although the pump is close to the group velocity dispersion minimum, this is not the case for the signal and idler photons. After a certain distance, the pump pulse no longer overlaps with the pair pulse, thus meaning that we cannot sum up coherently all the operators. Using the propagation constant determined in \sref{QAP}, we estimated the walk-off distance to be about $15\,cm$ (for our pulses). Therefore we have to apply our numerical calculation over this length and eventually to multiply accordingly to cover the actual length of the fibre.

From \eref{bdw} and \eref{bdw2} , we should expect the bandwidths to remain constant beyond this length. Note however, in \eref{Nsfinal}, the number of pairs created always grows linearly with the length. As a consequence, the walk-off distance will not play a role in the number of created photons and monitoring the bandwidth fluctuation versus fibre length is the only way we have to experimentally estimate the figure associated to the walk-off.

\section{Experiment\label{exp}}

\subsection{Setup}

\begin{figure}[h]
	\centerline{\scalebox{0.8}{\includegraphics{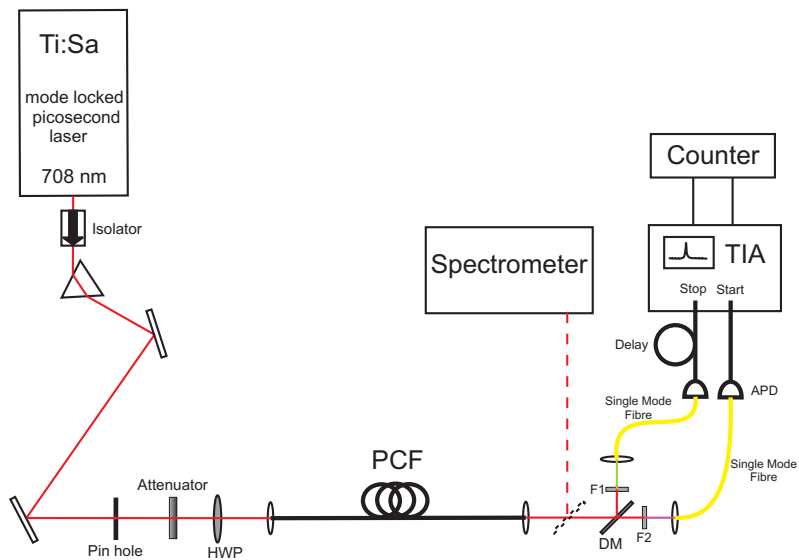}}}
	\caption{Optical layout. Laser, $708\,nm$ Ti:Sa laser; P, prism; HWP, halfwave plate; PCF, photonic crystal fibre; M, protected silver mirror ($R\leq 95\%$); DM, dichroic mirror (centered\@$700\,nm$, $T\leq85\%$, $R\leq90\%$); F1, $570\,nm$ band-pass filter, bandwidth $40\,nm$, $T=80\%$; F2, $880\,nm$ band-pass filter, bandwidth $40\,nm$, $T=80\%$; APD, Silicon single photon detector ($\eta_s\approx 60\%$ and $\eta_i\approx 33\%$).\label{manip}}
\end{figure}
In order to estimate the brightness of our source we used the coincidence setup depicted in \fref{manip} where a mode-locked picosecond Ti:Sapphire pump laser (Spectra Physics - Tsunami) emitting $\sim2\,ps$ pulses with a repetition rate of $80\,Mhz$ is sent, through an optical isolator, onto a prism P to remove in-band light from the pump laser spontaneous emission. A pin hole is then used to improve the pump mode and eventually several attenuators bring the power down so that up to $1\,mW$ average power is launched into the fibre. Since the PCF is birefringent and supports two modes, a half wave-plate (HWP) is used to align the pump polarization along one axis thus preventing polarization scrambling and creating pairs with the same polarization as the pump beam. The output of the fibre is collimated using an aspheric lens, followed by a dichroic mirror centered at $700\,nm$ to spread the incoming beam into two arms, one corresponding to the signal channel and the other to the idler, where band-pass filters F1 and F2 centered at $570\,nm$ and $880\,nm$ respectively (width $40\,nm$, $T>80\%$) are used to remove in-line pump and background light. Each photon of the pair is then launched into single mode fibres that are connected to two Silicon avalanche photodiodes (APD). The detected photons are counted in a dual-channel counter and the coincidences between the two APDs are analyzed using a time interval analysis system (TIA).

\subsection{Results}
Looking now at the coincidence test bench in order to determine the brightness of our source, we measured the number of single counts in both signal and idler channels, while we recorded the number of coincidences. This experimental protocol amounts to recording the percentage of detections in the ``start channel'' which have been stopped by detection in the ``stop channel'' in the following time interval and effectively provides a direct estimate of the ``lumped'' efficiency of the stop channel.
\begin{figure}[h]
	\centerline{\scalebox{1.2}{\includegraphics{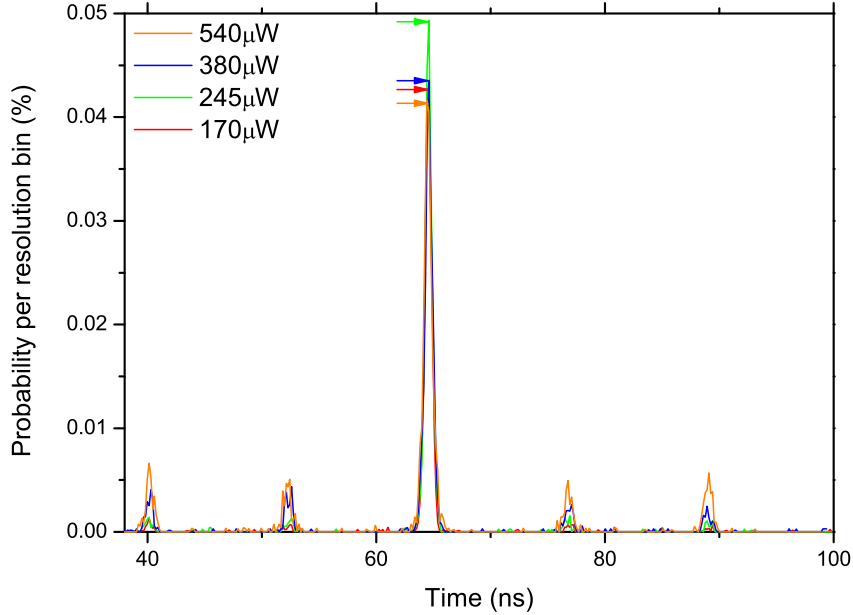}}}
	\caption{Time interval histogram showing the coincident photon detection peak and also a zoom on one of the accidental coincidence peak for different pump powers. The instrument displays the probability that a start pulse is stopped within a given time bin. Here the time-bin width is 156 ps. The time between two peaks reflects the pump laser repetition rate. However the width of the peaks is limited by the response time of the detectors which is typically hundreds of picoseconds (rather than the actual duration of the pump pulses).\label{coinc}}
\end{figure}

\subsubsection{Photon pair rate}

The central peak on \fref{coinc} corresponds to signal and idler photons belonging to the same pulse. The small satellite peaks stand for uncorrelated events, i.e. signal and idler coming from subsequent pulses, whether they are actual pairs of photons or background photons. It is interesting to note the satellite peaks grow with pump power, whereas the central peak remains constant (the arrows point at its height). The central peak reflects the lumped collection efficiency which is relatively constant with increasing pump power (confirming correlated photon pair creation), while the satellite ones are linked to the probability of having a photon from a different pair or a background photon in a neighbouring pulse. In the absence of background the satellite peaks would thus be indicative of the rate of generation of pairs of pairs, or 4-photon events. It is these events that could be useful for multi-photon interference experiments and quantum information applications. However, we are here interested in producing no more than one pair of photons per pulse and we still have to take into account both the background and the multi-photon pair rate in the central peak as they increase the photon count probability. Thus we will introduce $C_{raw}$ as the raw coincidence rate in the central peak and we will calculate the accidental coincidence rate thanks to the satellites peaks $C_b$. This allows us to write for $C_{raw}$:
\begin{equation}\label{C}
	C_{raw}=\mu_s\eta_s\mu_i\eta_i r + C_b
\end{equation}
where $r$ is the actual photon pair rate; $\eta_s$ and $\eta_i$ are the APD quantum efficiencies at $586\,nm$ and $894\,nm$, while the net optical transmission and launch efficiencies into single mode fibre of each arm are $\mu_s$ and  $\mu_i$. We defined the single counting rates in the signal and idler APDs as:
\begin{eqnarray}
	N_s^{raw}&=&\mu_s\eta_s r + \sum_{n=2}^{n=\infty}\mu_s\eta_s r^n + B_s\label{N}\\
	N_i^{raw}&=&\mu_i\eta_i r + \sum_{n=2}^{n=\infty}\mu_i\eta_i r^n + B_i\label{N2}
\end{eqnarray}
where $B_s$, $B_i$ are total background rates and $\mu_s\eta_s r^n$, $\mu_i\eta_i r^n$ stand for the multiphoton pair contributions. Here, we separated the background and multiphoton contribution in \eref{N} and \eref{N2}, as we can already consider the latter being negligible for low average powers. For instance, as long as we remain under 0.1 photon pairs per pulse, the high order contribution should remain a tenth of the total. On the other hand, the background might not be negligible and is mostly due to Raman scattering in our case~\cite{PCFUK}. Assuming we can single out this background contribution, as shown in reference~\cite{tanzEPJD}, we can now use the singles counting rates and the coincidence rates to estimate the actual rate of pairs $r$ created inside the PCF using this equation:
\begin{eqnarray}\label{actR}
	N_s^{raw}=\mu_s\eta_s r + B_s&&\nonumber\\
	N_i^{raw}=\mu_i\eta_i r + B_i&\qquad\Rightarrow& r=\frac{\left(N_s^{raw}-B_s\right)\left(N_i^{raw}-B_i\right)}{\left(C_{raw}-C_b\right)}\\
	C_{raw}=\mu_s\eta_s\mu_i\eta_i r + C_b&&\nonumber
\end{eqnarray}
As previously stated, $B_i$ and $B_s$ include the Raman background rates in the APDs and we have to single out these detections. The number of created photon pairs is proportional to the square of the peak intensity and the Raman scattering grows roughly linearly at these relatively low pump powers. Thus switching the laser from the pulsed regime to the CW regime reduces the photon pair rate to negligible levels while keeping the Raman and other sources of background constant for the same average power. We propose to validate the technique by recording the number of events in the two APDs versus the pump power for each case (i.e. pulsed and CW) and studying the evolution of the counting rate summarized in \tref{mesR}. Here a linear tendency would be the proof of spontaneous Raman scattering, while pure quadratic behavior would be the proof of actual photon pairs. All the plotted data have been corrected to compensate the non-linearity of the Si-APD for high counting rate following the datasheet~\cite{SPMAPD}.

\begin{table}[h]
\caption{Experimental measurement of the coincidence rate versus the pump power for $0.2\,m$ of PCF. The dark count rate in both APD is $\sim 400\,Hz$.\label{mesR}}
\begin{indented}
\lineup
\item[]\begin{tabular}{@{}*{7}{l}}
\br                              
Power&$N_s^{raw}$&[$N_s^{CW}$]&$N_i^{raw}$&[$N_i^{CW}$]&$C_{raw}$&[$C_b$]\cr 
\mr
$960\,\mu\!W$ & $2070\,kHz$&$[5.0\,kHz]$ & $917\,kHz$&$[54\,kHz]$  & $234\,kHz$&$[16\,kHz]$  \cr
  $660\,\mu\!W$ & $1031\,kHz$&$[3.0\,kHz]$ & $460\,kHz$&$[38\,kHz]$  & $115\,kHz$&$[3.8\,kHz]$  \cr
  $490\,\mu\!W$ & $601\,kHz$&$[2.0\,kHz]$ & $261\,kHz$&$[25\,kHz]$  & $66\,kHz$&$[1.4\,kHz]$  \cr
  $340\,\mu\!W$ & $298\,kHz$&$[1.2\,kHz]$ & $135\,kHz$&$[18\,kHz]$  & $33\,kHz$&$[0.1\,kHz]$  \cr
  $200\,\mu\!W$ & $107\,kHz$&$[0.9\,kHz]$ & $52\,kHz$&$[11\,kHz]$  & $12\,kHz$&$[0\,kHz]$  \cr
  $0\,\mu\!W$   & $0.4\,kHz$&$[0.4\,kHz]$ & $0.4\,kHz$&$[0.4\,kHz]$  & $0\,kHz$&$[0\,kHz]$  \cr 
\br
\end{tabular}
\end{indented}
\end{table}

\begin{figure}[h]
	\centerline{\scalebox{1.14}{\includegraphics{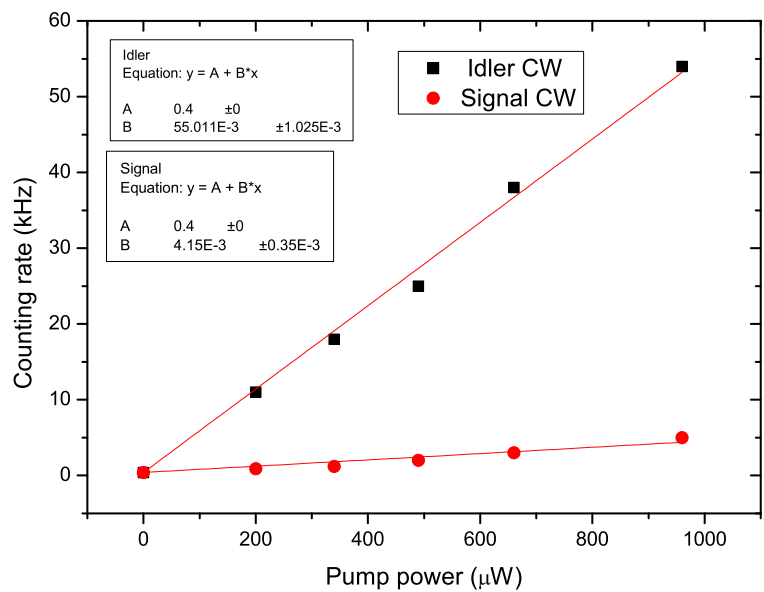}}}
	\caption{Corrected raw counting rate as function of the pump power in both APD in CW regime for $0.2\,m$ of fibre.\label{nsnicw}}
\end{figure}

We first plot in \fref{nsnicw} the number of detections versus the pump power in CW regime. We clearly see the negligible level of Raman scattering at the signal wavelength compared to the idler wavelength, as expected as the stokes Raman scattering is often considered negligible compared to the anti-stokes one. In both cases, the fit is purely linear (quadratic term is negligible) and being the proof of spontaneous Raman scattering process and possible leakage of pump light. We can therefore use these figures as a good estimation of the Raman background rate for any given mean pump power.
\begin{figure}[h]
	\centerline{\scalebox{1.14}{\includegraphics{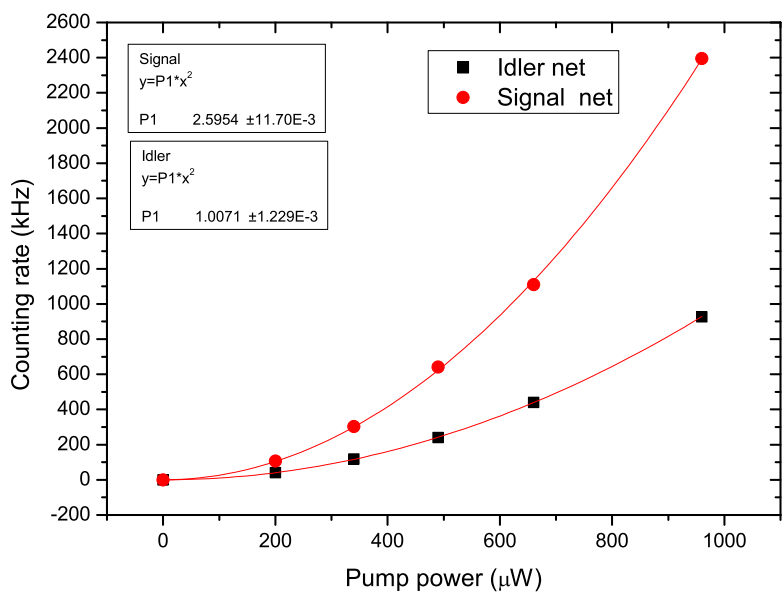}}}
	\caption{Corrected net counting rate as function of the pump power in both APD in pulsed regime for $0.2\,m$ of fibre. The net counting rate stands for the Raw counting rate minus the CW counting rate.\label{nsninet}}
\end{figure}

We then plot in \fref{nsninet} the number of net detections versus the pump power. We define the net detections as being the number of detections in pulsed regime minus the number of detections for the same average power in CW regime. We clearly see that the number of net detection versus the pump power is purely quadratic, thus highlighting that we are dealing with actual photon pairs. We can then now identify $B_s$ and $B_i$ as the CW counting rate and calculate in \tref{tableR} the actual number of photon generated in the fibre using \eref{actR}. To perform the calculation, we don't have to correct the counting rate from the APD, as $\mu_{s,i}$ can include all kinds of losses for the twin photons (even APD non linearity). If we eventually compare the counting rate associated with the actual Raman scattering to the counting rate associated with the actual photon pair, it confirms that the background in the signal channel is extremely low ($\leq0.3\%$ of the total rate) while in the infra-red we estimate the background rate to remain relatively low ($\leq6\%$).

\begin{table}[h]
\caption{Experimental estimation of the photon pair rate versus the pump power for $0.2\,m$ of PCF. From left to right, we first recall the associated pump power and the theoretical photon pair rate using \eref{Nsfinal}, followed by the experimental photon pair rate using \eref{actR}. For information purpose, we added the mean number of photon per pulse ($\frac{r_{exp}}{R_{laser}}$), the coincidence to accidental contrast ($\frac{C_{raw}-C_b}{C_b}$) and the lumped probability of detecting respectively the signal ($\mu_s\eta_s$) and idler photons ($\mu_i\eta_i$).\label{tableR}}
\begin{indented}
\lineup
\item[]\begin{tabular}{@{}*{7}{l}}
\br                              
Power & $r_{th}$ ($s^{-1}$) & $r_{exp}$ ($s^{-1}$) & $\left\langle n_s\right\rangle_{exp}$ &C/A& $\mu_s\eta_s$ & $\mu_i\eta_i$\cr 
\mr
 	$960\,\mu\!W$ & $\sim8.05\!\times\!10^6$ &$8.46\!\times\!10^6$& 0.11  &15:1& $0.235$ & $0.106$\cr
  $660\,\mu\!W$ & $\sim3.81\!\times\!10^6$ &$4.08\!\times\!10^6$& 0.05  &36:1& $0.240$ & $0.109$\cr
  $490\,\mu\!W$ & $\sim2.10\!\times\!10^6$ &$2.31\!\times\!10^6$& 0.03  &55:1& $0.244$ & $0.109$\cr
  $340\,\mu\!W$ & $\sim1.10\!\times\!10^6$ &$1.14\!\times\!10^6$& 0.015  &220:1& $0.244$ & $0.109$\cr
  $200\,\mu\!W$ & $\sim0.35\!\times\!10^6$ &$0.43\!\times\!10^6$& 0.006  &--& $0.223$ & $0.109$\cr
  \br
\end{tabular}
\end{indented}
\end{table}

If we compare the photon pair rate to the theoretical rate, we have here good agreement. However, one has to bear in mind that the estimation of the mode area and the peak power could lead to large error in the expected pair rate $r_{th}$. We found it useful to recall the lumped probability to detect the photons of the pairs. These important experimental parameters are useful for evaluating the possibility to perform multi-photon pairs experiment as discussed in \sref{disc}. Here, provided a detection efficiency $\eta_s\approx0.60$ and $\eta_i\approx0.33$~\cite{SPMAPD}, we found a global coupling efficiency from the PCF fibre to the single mode fibre of $\mu_s\approx0.42$ and $\mu_i\approx0.30$. These figures include the fibre coupling, filters, dichroic mirror and lenses losses for signal and idler channels respectively. By correcting these results with the filters and dichroic mirror losses, we find a coupling efficiency into single mode fibre of respectively $0.58$ and $0.44$ for the signal and idler photons. We also found useful to introduce the coincidence to accidental contrast. This figure of merit has been introduced in~\cite{PCFUS2} to quantify the quality of any presented coincidence rate. Our results clearly illustrate that this measure can be misleading as the coincidence to accidentals ratio drops with increasing pump power. This is due primarily to the increasing number of overlaps between pairs of pairs, which we hope to exploit in future multiphoton interference experiments.

\subsubsection{Walk-off}

Taking into account the walk-off distance we identified in \sref{limit}, we should expect the bandwidth to be constant versus the fibre length whenever the fibre is longer than $15\,cm$. For several fibre length, we recorded the pump bandwidth, measured the signal fluorescence bandwidth and compared it to \eref{bdw}.
\begin{figure}[h]
	\centerline{\scalebox{1.1}{\includegraphics{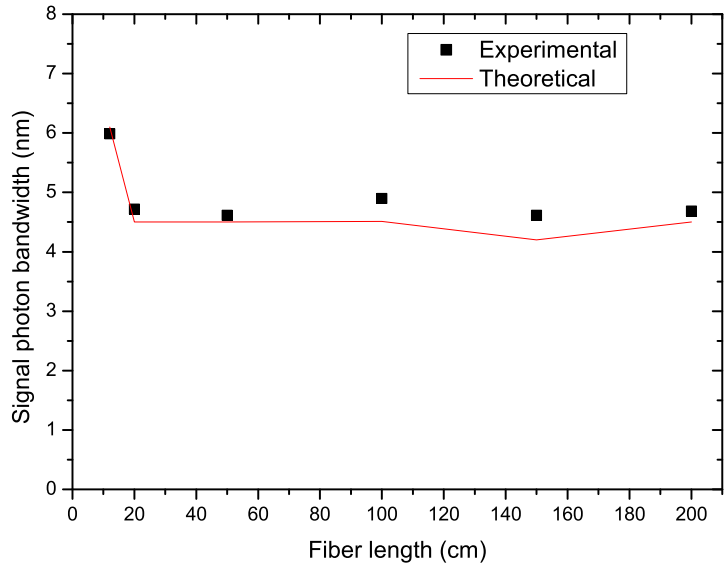}}}
	\caption{Comparison of the measured and predicted bandwidth of the emitted signal photons.\label{bwplot}}
\end{figure}
From \fref{bwplot}, one can note the quite good matching between the measured and predicted bandwidth and the similar behavior of both bandwidths versus the fibre length. As predicted, the measured bandwidth remains constant whenever the fibre is longer than the walk-off distance which we estimate to be around $15\,cm$. More measurements around $15\,cm$ would be needed to reduce the uncertainty about the walk-off experimental determination.

\section{Discussion\label{disc}}

This source has to be usable for future quantum interference experiments involving three or more photons created as two or more pairs~\cite{fundprob}. We dedicated the following section to estimate the potential of this source. To do so, we used realistic requirements for interference effects between separate pair-photons and attached experimental figures to a typical four-photon quantum experiment.

\subsection{Multi-photon pair weight in satellites peaks in \fref{coinc}}

We showed in the previous sections that the spontaneous Raman scattering was negligible compared to the photon pair rate. However this background may be equivalent in intensity to the probability of creating two pairs of photons. We can use the diagram in \fref{coinc} to state that our source exhibits a background low enough to allows multi-photon pairs experiments. The spontaneous Raman scattering rate grows linearly with pump power whereas the photon pair rate grows quadratically. Hence, in \fref{coinc}, we would therefore expect the satellite peaks to decrease in significance as we increase pump power as the ratio $\mbox{Raman}/\mbox{photon pair}$ decreases.  Experimentally, the satellite peaks in \fref{coinc} increase clearly showing that the background coincidence rate at high counting rates is dominated by random overlap of more than one pair of photons. The satellites peaks are thus linked to the multi-photon pair probability, they are indicative of the four photon coincidence rate we will get in future experiments and are variables we want to maximize.

For these multi-photon quantum information experiments, we require a narrower bandwidth so that the coherence length is equivalent to the pulse length~\cite{fundprob}. A quantum interference experiment involving four-fold coincidence between photons coming from two separated sources would require a filter of order $0.2\,nm$ bandwidth in the green ($0.4\,nm$ in the IR). Such filters will transmit only $40\%$ of in-band light thus halving our effective efficiencies and collect only $\sim 1/25$ of the available spectrum. However our present coincidence rates are limited by the detector saturation. Here the singles counting rates would be significantly reduced thus allowing an increase in pump power. Using our source with a pump power of $\sim6\,mW$, the expected rate of photon pairs detected within this bandwidth is $\sim 9\!\times\!10^4\,s^{-1}$, which means a rate of four photon events $\sim100\,s^{-1}$, two orders of magnitude higher than any previous experiment. It is also important to underline, that our Raman background rate in the APDs, will be reduced by a bandwidth factor of $\sim 1/100$ due to its broadband nature thus improving even more the performances of the source.
\section{Conclusion\label{CCL}}

We have presented in this article both theoretical and experimental aspects of photon pair generation using four wave mixing in a photonic crystal fibre. We have clearly demonstrated a good agreement between the theory and experiment. Any realistic multi-photon quantum experiment requires a precise timing using ultra-short pulses, thus we have developed the first quantum model suitable for such a regime. We used it to suggest a numerical estimation of the photon pair rate from a photonic crystal fibre provided a good knowledge of the mode propagating in the fibre. We have validated the model by comparison with the experimental measurement of picosecond-pulsed photon pairs generated by four-wave mixing in a single-mode optical fibre. The source we used is polarized, bright, narrowband, single-mode and tunable by varying laser wavelength or fibre parameters. The wide separation of the generated pair wavelengths means that most of the background can be avoided. All these advantages make this new source of photon pairs more appropriated compared to conventional ones for quantum information processing applications.

\ack WJW is a Royal Society University Research Fellow. The work is partly funded by UK EPSRC (QIP IRC and 1-phot), EU IST-2001-38864 RAMBOQ and FP6-2002-IST-1-506813 SECOQC.

\appendix

\section{Derivation of the interaction Hamiltonian in the pulsed case\label{hamiltonian}}
The interaction Hamiltonian in a volume within the fibre is:
\begin{equation}
H_{int}=\int_VUdV
\end{equation}
and considering the energy density associated with four fields coupled in a third order non-linear medium to be:
\begin{equation}
U=\epsilon_0 \chi^{(3)}(\Omega_p,\omega_s,\omega_i)E_p^2 E_s E_i
\end{equation}
In the four wave mixing-process, we will look at two pump frequency components $\left\{\omega_{p1},\omega_{p2}\right\}$ interacting one with each other.
This new component of our interest is a Gaussian defined as follow where $G_p\left(\omega\right)=e^{-\frac{\omega^2\sigma^2}{2}}$ is the weight of one frequency component of the pump pulse:
\begin{eqnarray}\label{Ep2}
\fl\mathcal{E}_p(\omega_{p1})\times\mathcal{E}_p(\omega_{p2})=\frac{E_{p_o}^2}{4}&\times & G_p\left(\omega_{p1}\right)\left(e^{-i[\left(\Omega_p+\omega_{p1}\right)t-(k_p-\gamma P_p)z]}+h.c.\right)e_{p1}(\vec{r})\nonumber\\
&\times & G_p\left(\omega_{p2}\right)\left(e^{-i[\left(\Omega_p+\omega_{p2}\right)t-(k_p-\gamma P_p)z]}+h.c.\right)e_{p2}(\vec{r})
\end{eqnarray}
where $\frac{E_{p_o}^2}{4}$ is proportional to the density of energy per frequency units squared, and $e_p(\vec{r})$ is the transverse spatial variation with normalisation $\int\left|e_p(\vec{r})\right|^2 rdrd\theta=1$. We take the quantified fields for the signal ($l=s$) and idler ($l=i$) modes as:
\begin{equation}
E_l=\sqrt{\frac{\hbar\omega_l}{2\epsilon_l}}\frac{1}{\sqrt\mathcal{L}}\sum_{k_l}\left(a^\dag_le^{-ik_lz}-a_le^{ik_lz}\right)e_l(\vec{r})
\end{equation}
The four-wave mixing interaction only occurs if the energy conservation and phase-matching condition are satisfied:
\begin{eqnarray}
	\left(2\Omega_{p}+\omega_{p1}+\omega_{p1}\right)-\omega_s-\omega_i&=&0\label{en}\\
	\left(\vec{k}_{p1}+\vec{k}_{p2}\right)-\vec{k}_s-\vec{k}_i-2\gamma P_p\cdot \vec{u}_p&=&\vec{0}\label{ph}
\end{eqnarray}
where $\left\{\omega_{p1},\omega_{p2}\right\}$ are the relative frequencies around the central pump $\Omega_p$. Since we are in a collinear guided configuration, we can omit the vectorial terms by projecting on the $z$-axis. We finally get for eqs~\ref{en} and~\ref{ph}:
\begin{eqnarray}
	\left(2\Omega_{p}+\omega_{p1}+\omega_{p1}\right)-\omega_s-\omega_i&=&0\\
	\left(k_{p1}+k_{p2}\right)-k_s-k_i-2\gamma P_p&=&\vec{0}
\end{eqnarray}
where $k_{i,s,p}$ are the wave-vectors (propagation constants) of the idler, signal and pump photons; $P_p$ is the peak pump power and $\gamma$ is the nonlinear coefficient of the fibre
\begin{equation}
	\gamma=\frac{2\pi n_2}{\lambda_p A_{\mbox{\emph{eff}}}}
\end{equation}
where $n_2\approx2\times10^{-20}\,m^2/W$  is  the nonlinear refractive index of silica and $A_{\mbox{\emph{eff}}}$ is the effective area of the fibre mode. We then restrict the interaction Hamiltonian for one frequency component of the pump pulse to the phase matching situation when $\Delta k\approx 0$. This approximation is justified since all the other combinations (i.e. $\left[2k_p+k_s-k_i-2\gamma P,\ldots,2k_p+k_s+k_i-2\gamma P\right]$) of fields will lead to null terms after integration over the length of the fibre.
\begin{equation}
\fl\mathcal{H}_{int}=SI\hbar \int_{-L/2}^{L/2}\frac{1}{\mathcal{L}}\sum_{k_s,k_i}G_p\left(\omega_{p1}\right)G_p\left(\omega_{p2}\right)
\left(a_s^\dag a_i^\dag e^{-i\left(2\Omega_p+\omega_{p1}+\omega_{p2}\right)t}e^{i\Delta kz}+h.c.\right)dz\nonumber
\end{equation}
where $I=\int\!\!\!\int e_{p1}(\vec{r})e_{p2}(\vec{r})e_s(\vec{r})e_i(\vec{r}) dr d\theta$ is a normalised factor and $\Delta k = k_{p1}+k_{p2}-k_s-k_i-2\gamma P_p$ the phase matching coefficient. It is convenient to define the gain parameter as:
\begin{equation}\label{SA}
S=\epsilon_0 \chi^{(3)}\frac{E_{p_o}^2}{4}\sqrt{\frac{\omega_s\omega_i}{4\epsilon_s\epsilon_i}}
\end{equation}
By carrying out the integral over the length of the fibre, we get the following interaction Hamiltonian:
\begin{equation}
\fl\mathcal{H}_{int}=SI\hbar\frac{1}{\mathcal{L}}\sum_{k_s,k_i}G_p\left(\omega_{p1}\right)G_p\left(\omega_{p2}\right)
\left(a_s^\dag a_i^\dag e^{-i\left(2\Omega_p+\omega_{p1}+\omega_{p2}\right)t}L sinc\left(\frac{\Delta kL}{2}\right)+h.c.\right)
\end{equation}
It is important to note that the Hamiltonian $\mathcal{H}_{int}$ only stands for a frequency component of the pulse. Therefore we will have to integrate later over the whole pump pulse frequencies.

\section{The Heinsenberg equation of motion\label{eqmotion}}

In the standard Heisenberg representation the evolution of any operator $A$ is defined by
\begin{equation}\label{heisenA}
\frac{dA}{dt}=-\frac{i}{\hbar}\left[A,H_{tot}\right]
\end{equation}
In the case of $A=a^\dag_l$, equation \ref{heisenA} gives
\begin{equation}\label{asA}
	\fl\frac{da^\dag_s}{dt}=i\omega_sa^\dag_s-i\frac{SIL}{\sqrt{\mathcal{L}}}\sum_{k_i}G_p\left(\omega_{p1}\right)
	G_p\left(\omega_{p2}\right)\left(a_i e^{-i\left(2\Omega_p+\omega_{p1}+\omega_{p2}\right)t}sinc\left(\frac{\Delta kL}{2}\right)\right)
\end{equation}
We now assume that the gain $S$ is small and that we can make the slowly varying envelope approximation (SVEA) $a_l(t)=\tilde{a}_l(t)e^{-i\omega_lt}$ where  $l=s,i$ leading to simplify eq~\ref{asA} into:
\begin{equation}	
\frac{d\tilde{a}^\dag_s}{dt}=-i\frac{SIL}{\sqrt{\mathcal{L}}} \sum_{k_i}G_p\left(\omega_{p1}\right)G_p\left(\omega_{p2}\right)
\left(\tilde{a}_i e^{i\Delta\omega t}sinc\left(\frac{\Delta kL}{2}\right)\right)
\end{equation}
where $\Delta\omega = \left(2\Omega_p+\omega_{p1}+\omega_{p2}\right)-\omega_s-\omega_i$.
Since $\tilde{a}_i(t)$ is slowly varying, we can assume that $\tilde{a}_i(t)\approx\tilde{a}_i(0)$ and for each Fourier components of the pulse, we can easily integrate over the time from $0$ to $t\rightarrow\infty$ (we assumed that the system --- nonlinear medium and pump laser --- has been ready for a long time) leading to the $\delta$-function:
\begin{equation}	
\fl\tilde{a}^\dag_s(t\rightarrow\infty)=\tilde{a}^\dag_s(0)-i\frac{SIL}{2\sqrt{\mathcal{L}}}\sum_{k_i}G_p\left(\omega_{p1}\right)
G_p\left(\omega_{p2}\right)\left(\tilde{a}_i(0) \delta(\Delta\omega)sinc\left(\frac{\Delta kL}{2}\right)\right)
\end{equation}
It is interesting to note here that the energy conservation lies in the $\delta$-function of $\Delta\omega$, as it would be in the monochromatic pump case. If we now sum up over the whole pump frequencies, we get:
\begin{eqnarray}\label{operatorA}
\fl\tilde{a}^\dag_s(t\rightarrow\infty)&=&\tilde{a}^\dag_s(0)-i\frac{SIL}{2\sqrt{\mathcal{L}}}\times\\
&&\int\!\!\! \int\sum_{k_i}G_p\left(\omega_{p1}\right)G_p\left(\omega_{p2}\right)
\tilde{a}_i(0) \delta(\Delta\omega)sinc\left(\frac{\Delta kL}{2}\right)\,d\omega_{p1} d\omega_{p2}\nonumber
\end{eqnarray}
It is possible to invert the integrals over $\left\{d\omega_{p1},d\omega_{p2}\right\}$ and the sum over $k_i$ as they are independent variables. This leads us to integrate over a monochromatic field operator $\tilde{a}^\dag_s$, for a given (monochromatic) mode in the idler. As a consequence the dirac function helps to compute the double integral since $d\omega_{p1}$ has to move as the opposite of $d\omega_{p2}$ to keep satisfied the energy conservation. Given the basic energy conservation $2\Omega_p=\omega_{s_0}+\omega_{i_0}$, we are here performing the integration for a signal mode $k_s$ and an idler mode $k_i$ satisfying the energy conservation
\begin{equation}\label{newenA}	\left(\Omega_p+\omega_{p1}\right)+\left(\Omega_p+\omega_{p2}\right)=\left(\omega_{s_0}+\Delta\omega_{k_{s}}\right)+\left(\omega_{i_0}+\Delta\omega_{k_{i}}\right)
\end{equation}
where we have introduced:
\begin{itemize}
	\item $\Delta\omega_{k_{s}}$ --- the frequency shift for the signal mode compared to the central one $\omega_{s_0}$
	\item $\Delta\omega_{k_{i}}$ --- the frequency shift for the idler mode compared to the central one $\omega_{i_0}$	
\end{itemize}
For instance, \eref{newenA} requires $\omega_{p2}$ to become $\omega_{p2}-\Delta\omega_{k_i}$ (while $\omega_{p1}$ becomes $\omega_{p1}-\Delta\omega_{k_s}$) to satisfy the energy conservation. Eq~\ref{operatorA} becomes then:
\begin{eqnarray}
\fl\tilde{a}^\dag_s(t\rightarrow\infty)&=&\tilde{a}^\dag_s(0)-i\frac{SIL}{2\sqrt{\mathcal{L}}}\times\\
&&\sum_{k_i}\int\!\!\!\int e^{-\frac{\left(\omega_{p1}-\Delta\omega_{k_s}\right)^2\sigma^2}{2}}e^{-\frac{\left(\omega_{p2}-\Delta\omega_{k_i}\right)^2\sigma^2}{2}}\tilde{a}_i(0) \delta(\Delta\omega)sinc\left(\frac{\Delta kL}{2}\right)\,d\omega_{p1} d\omega_{p2}\nonumber
\end{eqnarray}
The $\delta$-function is used to reduce the number of integrals, by changing $\omega_{p2}$ into $-\omega_{p1}$. We can also introduce the new variable $\omega_{p1}\rightarrow\omega'_1-\frac{\Delta\omega_{k_s}-\Delta\omega_{k_i}}{2}$ and assume that $sinc\left(\frac{\Delta kL}{2}\right)$ will not change significantly while we scan $\omega_{p1}$ over $\Delta\omega_p$. We now find after integration over $\omega'_1$ from $-\infty$ to $\infty$:
\begin{equation}\label{asfA}
\tilde{a}^\dag_s(t\rightarrow\infty)=\tilde{a}^\dag_s(0)-i\frac{SIL}{2\sqrt{\mathcal{L}}}\sum_{k_i} \frac{\sqrt{\pi}}{\sigma} e^{-\frac{\left[\Delta\omega_{k_i}+\Delta\omega_{k_s}\right]^2\sigma^2}{4}}\tilde{a}_i(0)sinc\left(\frac{\Delta kL}{2}\right)
\end{equation}
In \eref{operatorA}, we were considering a monochromatic component of the pump pulse. Thus the theory still involved a Dirac delta function arising from the energy conservation. However, in \eref{asfA} we have left the monochromatic pump regime and lost the Dirac function while integrating over the pump pulse. However it is interesting to note that we still have energy conservation in the broadband pump regime represented within the Gaussian function with width limited by the bandwidth of our pump pulse.
\noindent In order to calculate the mean number of photons $\left\langle N_s\right\rangle$ over all the possible modes $k_s$, we define the operator $W_s\left(t\right)$ to be given by a sum over the $k_s$: 
\begin{equation}\label{WsA}
	W_s\left(t\right)=\frac{1}{\sqrt{\mathcal{L}}}\sum_{k_s}\tilde{a}_s(t)
\end{equation}
so that the mean number of photon per pulse is given by:
\begin{equation}\label{NsA}
		\left\langle N_s\right\rangle = _i\bra {0}_s\bra {0}W^\dag_sW_s\ket {0}_s\ket {0}_i
\end{equation}
From Eqs.~\ref{asfA} and~\ref{WsA} we found for $\left\langle N_s\right\rangle$ the following equation
\begin{eqnarray}
\fl\left\langle N_s\right\rangle & = &  \left(\frac{SIL}{2\mathcal{L}}\right)^2\left(\frac{\sqrt{\pi}}{\sigma}\right)^2\sum_{k_s,k_i}\sum_{k_{s'},k_{i'}} \Bigg[ e^{-\frac{\left[\Delta\omega_{k_i}+\Delta\omega_{k_s}\right]^2\sigma^2}{4}}sinc\left(\frac{\Delta kL}{2}\right)\times\\
&& \qquad\qquad\qquad\qquad e^{-\frac{\left[\Delta\omega_{k_{i'}}+\Delta\omega_{k_{s'}}\right]^2\sigma^2}{4}}sinc\left(\frac{\Delta k'L}{2}\right)\Bigg]\, _i\bra {0}_s\bra {0}\tilde{a}_i\tilde{a}_{i'}^\dag\ket {0}_s\ket {0}_i\nonumber
\end{eqnarray}
The commutation properties imply $k_{s'}=k_s$, $k_{i'}=k_i$ and $\Delta k'=\Delta k$ and allow us to simplify the formula
\begin{equation}
\left\langle N_s\right\rangle = \left(\frac{SIL}{2\mathcal{L}}\right)^2\left(\frac{\sqrt{\pi}}{\sigma}\right)^2
\sum_{k_s,k_i}\left[sinc^2\left(\frac{\Delta kL}{2}\right)e^{-\frac{\left[\Delta\omega_{k_i}+\Delta\omega_{k_s}\right]^2\sigma^2}{2}}\right]
\end{equation}
In the limit $\mathcal{L}\rightarrow\infty$, the sum over $k_{s,i}$ can be replaced by an integral in the usual way
\begin{equation}
	\left\langle N_s\right\rangle = \left(\frac{SIL}{2}\right)^2\left(\frac{\sqrt{\pi}}{\sigma}\right)^2
 \int\!\!\! \int
sinc^2\left(\frac{\Delta kL}{2}\right)e^{-\frac{\left[\Delta\omega_{k_s}+\Delta\omega_{k_i}\right]^2\sigma^2}{2}}dk_sdk_i
\end{equation}
As a final step, we can now convert the integrals over modes into integrals over frequencies~\cite{klein} where $v_{g_s}$ and $v_{g_i}$ are the group velocity of signal and idler wave in the fibre.
\begin{equation}\label{nsfa}
\fl\left\langle N_s\right\rangle = \left(\frac{SIL}{2}\right)^2\left(\frac{\sqrt{\pi}}{\sigma}\right)^2
\int\!\!\!\int\!
sinc^2\left(\frac{\Delta kL}{2}\right)e^{-\frac{\left[\Delta\omega_{k_s}+\Delta\omega_{k_i}\right]^2\sigma^2}{2}}\frac{d\Delta\omega_{k_s}}{v_{g_s}}\frac{d\Delta\omega_{k_i}}{v_{g_i}}
\end{equation}

\section{Estimation of the number of photons\label{acalc}}

This integral can be numerically evaluated if we now develop the $\Delta k$-function around the phase matched frequencies using first order Taylor power series:
\begin{equation}
	\Delta k=-\mathcal{N}_p\left(d\omega_{p1}+d\omega_{p2}\right)
	+\mathcal{N}_s d\omega_s
	+\mathcal{N}_i d\omega_i
\end{equation}
where $\mathcal{N}_l=\left[\omega_l\left.\frac{\partial n_l}{\partial\omega}\right|_{\omega_l}+n_l\right]$. Then with the help of energy conservation  ($-d\omega_{p1}-d\omega_{p2}+d\omega_i+d\omega_s=0$) we find:
\begin{eqnarray}
	\Delta k&=&\frac{\left(\mathcal{N}_s-\mathcal{N}_p\right)}{c}d\omega_s+\frac{\left(\mathcal{N}_i-\mathcal{N}_p\right)}{c}d\omega_i
\end{eqnarray}
We can now write Eq.\ref{nsfa}:
\begin{eqnarray}~\label{nsfa2}
	\left\langle N_s\right\rangle&=&\left(\frac{SIL}{2}\right)^2\left(\frac{\sqrt{\pi}}{\sigma}\right)^2\times\\
	&&\fl\int\!\!\! \int 
sinc^2\left(\frac{\left(\mathcal{N}_s-\mathcal{N}_p\right)L}{2c}\Delta\omega_{k_s}+\frac{\left(\mathcal{N}_i-\mathcal{N}_p\right)L}{2c}\Delta\omega_{k_i}\right)e^{-\frac{\left[\sigma\Delta\omega_{k_s}+\sigma\Delta\omega_{k_i}\right]^2}{2}}\frac{d\omega_s}{v_{g_s}}\frac{d\omega_i}{v_{g_i}}\nonumber
\end{eqnarray}
It is quite easy to integrate this product making the following change of variables ($X=\frac{\Delta\omega_{k_s}+\Delta\omega_{k_i}}{2}$ and $Y=\frac{\Delta\omega_{k_s}-\Delta\omega_{k_i}}{2}$). Then it becomes obvious that if we perform the Y integral first and consider the $sinc$ function as a gate of height 1 and FWHM=$\frac{\left(\mathcal{N}_s-\mathcal{N}_i\right)L}{2c}\Delta Y=\pi$, the system is then easy to solve and we find:
\begin{equation}
	\left\langle N_s\right\rangle = \left(\frac{SIL}{2}\right)^2\left(\frac{\sqrt{\pi}}{\sigma}\right)^2 \left(\frac{2\pi c }{\left(\mathcal{N}_s-\mathcal{N}_i\right)L}\right)\left(\frac{\sqrt{2\pi}}{2\sigma}\right)\frac{4}{v_{g_s}v_{g_i}}
\end{equation}
\begin{equation}
	\left\langle N_s\right\rangle = \left(\frac{SIL}{2}\right)^2\left(\frac{\pi\Delta\omega_p^2}{4\ln(4)}\right)^{3/2}\left(\frac{4\sqrt{2}\pi c}{\left(\mathcal{N}_s-\mathcal{N}_i\right)L}\right)\frac{1}{v_{g_s}v_{g_i}}
\end{equation}

\section*{bibliography}
\bibliography{Refs/babolive}

\end{document}